\documentclass[aps, epsf, amsfonts]{iopart}
\usepackage{graphics}
\usepackage{graphicx}
\usepackage{epsfig}
\begin{document}
\input epsf.tex

\title{An Overview of LISA Data Analysis Algorithms.}
\author{Edward K. Porter}
\address{Laboratoire AstroParticules et Cosmologie (APC)\\Universit\'e Paris 7 - Denis Diderot\\10, Rue Alice Domon \& Leonie Duquet\\75205 Paris Cedex 13\\France}
\vspace{1cm}
\begin{abstract}
\noindent The development of search algorithms for gravitational wave sources in the LISA data stream is currently a very active area of research.  It has become clear that not only does difficulty lie in searching for the individual sources, but in the case of galactic binaries, evaluating the fidelity of resolved sources also turns out to be a major challenge in itself.  In this article we review the current status of developed algorithms for galactic binary, non-spinning supermassive black hole binary and extreme mass ratio inspiral sources.  While covering the vast majority of algorithms, we will highlight those that represent the state of the art in terms of speed and accuracy.
\end{abstract}

\maketitle

\section{Introduction}
The ESA-NASA Laser Interferometer Space Antenna (LISA)~\cite{Ben98} will be able to detect the emitted gravitational waves (GWs) from a number of sources.  Amongst the strongest sources will be the inspiral and merger of comparable mass supermassive black holes, the inspiral of stellar mass black holes or neutron stars into central massive black holes and the galaxy containing tens of millions of quasi-monochromatic white dwarf binary systems.  The LISA detector is composed of three spacecraft which form an equatorial triangle and will work in the frequency range $10^{-5} \leq f/Hz\leq 1$.  The center of mass of the constellation traces out a circular orbit around the Sun at a distance of 1 AU and lies about $20^{o}$ behind the Earth.  The three spacecraft cartwheel in retrograde motion as they move around the Sun with a period of one year.  This motion induces amplitude, frequency and phase modulations in the gravitational wave signal.  The amplitude modulation is caused by the antenna pattern being moved across the sky.  Because LISA can be thought of as two separate detectors, measuring different polarizations of the GW, the phase modulations are cause by combinations of the polarizations.  Finally, the frequency or Doppler modulations are caused by the motion of the detector with respect to the source. 

While a lot of algorithms have been developed and tested on in-house data, a more communal way of algorithm testing exists.  The Mock LISA Data Challenges (MLDC\footnote[1]{http://www.tapir.caltech.edu/dokuwiki/listwg1b:home}) were initiated by the LISA International Science Team (LIST) at the end of 2005.  A taskforce exists which decides on the severity of the challenge, the types and number of sources etc. and also conducts the analysis of entries at the end of each challenge.  A challenge data set is regularly issued with a deadline ranging from six to twelve months.  These challenges are open to everyone within the GW community, and allow the community to simulate a realistic data analysis effort where the number of input sources and parameters are relatively unknown.

\section{Galactic Binaries.}
Binary systems consisting of pairs of white dwarves are expected to be one of the major sources of GWs in the LISA detector.  While we expect may tens of millions of such systems in the galaxy, only approximately 25,000 of these sources will actually resolvable~\cite{TRC06}.  The other sources remain in the data stream as an extra source of noise called {\em confusion noise}.  While detection of the so-called verification binaries will produce a good metric of our ability to conduct parameter estimation using gravitational waves, it is also imperative that we are able to successfully and confidently remove as many galactic binaries as is possible from the data stream.   It is possible that an untreated galaxy of galactic binaries could effect the success of parameter estimation for supermassive black hole binaries (SMBHBs) and even the detection of extreme mass ratio inspirals (EMRIs), as in general, these sources will lie below the noise floor that includes the galaxy.

The GWs from a galactic binary are described by an eight parameter set $\vec{\lambda} = \{A_0, \iota, \psi, \varphi_0, f_0, \dot{f_0}, \theta, \phi\}$, where $A_0$ is a constant amplitude, $\iota$ is the inclination of the orbital plane of the binary, $\psi$ is the GW polarization, $\varphi_0$ is the initial phase of the GW, $(f_0, \dot{f_0})$ are the monochromatic frequency and first derivative of the frequency and $(\theta, \phi)$ are the sky coordinates for the source.  Each polarization of the GW is characterized by \begin{eqnarray}
h_+ & = & A_0\left(1+\cos^2 \iota\right)\cos(\varphi_0 + \Phi(t)), \\  h_{\times} &=& -2 A_0 \cos\iota\sin(\varphi_0 + \Phi(t)),
\end{eqnarray}
where the phase term is described by
\begin{equation}
\Phi(t) = 2\pi f_0 t + \pi \dot{f_0}t^2 + 2\pi\left(f_0+\dot{f_0}\right)R_{\oplus}\sin\theta\cos\left(2\pi f_m t-\phi\right).
\end{equation}
The final term in the above expression describes the Doppler modulation to the phase due to the motion of the LISA detector about the sun.  The quantity $R_{\oplus}$ corresponds to the light crossing time for one AU (i.e. $\sim$500 seconds) and $f_m$=1/year  is the LISA modulation frequency.  The frequency derivative is described (to leading order) by
\begin{equation}
\dot{f_0} = \frac{96}{5} \pi^{8/3}f_{0}^{11/3}M_{c}^{5/3},
\end{equation}
where $M_c = m\eta^{3/5}$ is the chirp mass, $m=m_1+m_2$ is the total binary mass and $\eta=m_1 m_2 / m^2$ is the symmetric reduced mass ratio.  In detatched binaries the frequency derivative is only resolvable in the higher frequency sources.  Because of this, most galactic binary systems are sufficiently described by a seven parameter set.

\subsection{Search Algorithms for Galactic Binaries.}
It was shown quite early on that LISA data analysis would need the development of new search techniques.  The most commonly used algorithm within the ground based community is the construction of a template bank~\cite{Owen96,OS99}.  This is a grid of theoretical waveforms, or templates, placed in the parameter space.  Each intersection of the grid corresponds to a certain combination of parameters and represents a waveform solution.  In general, the templates are placed such that the minimal match between a template and a source achieves a preset threshold.  The problem with standard template banks is the number of templates needed scales geometrically with the number of parameter dimensions.  Once we have more than two dimensions, the number of templates quickly blows up.  It was shown by Cornish \& Porter~\cite{CP05} that using an F-Statistic~\cite{JKS98} maximised template grid, the search for a galactic binary assuming a minimal match of 97\% required between $10^6$ and $10^{10}$ templates, depending on the monochromatic frequency.  This study also showed that the scaling of template number with frequency changed from $f_{0}^{1.25}$ to $f_{0}^{5.88}$ beyond a frequency of $f_{0} = 1.6\times10^{-3} $ Hz due to a change in dimensionality of the problem,  as $\dot{f_0}$ now becomes resolvable.
  As the F-Statistic will be mentioned a lot in this article, it deserves further explanation.  The F-Statistic allows us to write the detector response in a single channel
\begin{equation}
h(t) = h_+(t)F^{+}(t) + h_{\times}(t)F^{\times}(t),
\end{equation}
in the form
\begin{equation}
h(t) = \sum_{i=1}^{4}a_{i}\left(\iota, \psi, A_0, \varphi_{0}\right)A^{i}\left(t;f_0, \dot{f}_0, \theta,\phi\right),
\label{eqn:strainFS}
\end{equation}
where the response is now composed of four quantities $a_i$ which are a function of the extrinsic parameters (i.e. those parameters that are dependent of the detector) and four time dependent $A^i$'s that are a function of the intrinsic parameters (i.e. parameters that describe the dynamics of the system).  Now assuming the detector output is a combination of a signal plus noise, i.e. $s(t)=h(t)+n(t)$, the noise weighted inner product between the data $s(t)$ and a template $h(t)$ is written
\begin{equation}\label{eqn:scalarprod}
\left<h\left|s\right.\right> =2\int_{0}^{\infty}\frac{df}{S_{n}(f)}\,\left[ \tilde{h}(f)\tilde{s}^{*}(f) +  \tilde{h}^{*}(f)\tilde{s}(f) \right],
\end{equation}
where
\begin{equation}
\tilde{h}(f) = \int_{-\infty}^{\infty}\, dt\, h(t)e^{2\pi\imath ft},
\end{equation}
is the Fourier transform of the time domain waveform $h(t)$ and $S_n(f)$ is the one sided noise power spectral density.   Now defining four constants $N^{i}=\left<s \left| A^{i}\right>\right.$, we can find a solution for the $a_{i}$'s in the form
\begin{equation}
a_{i} = M_{ij}N^{j},
\label{eqn:ais}
\end{equation}
where the M-Matrix is defined by
\begin{equation}
M_{ij} = \left(M^{ij}\right)^{-1} = \left<A^{i} \left| A^{j}\right>^{-1}\right. .
\end{equation}
We can now write the F-statistic as
\begin{equation}
{\mathcal F} = \frac{1}{2}M_{ij}N^{i}N^{j},
\end{equation}
which automatically maximizes the log-likelihood over the extrinsic parameters and reduces the search space to the sub-space of intrinsic parameters.  Once we have numerical solutions for the four $a_i$'s, we can  analytically maximize over the extrinsic parameters.  Furthermore, the F-Statistic is related to the SNR by ${\mathcal F} \approx SNR^2 / 2$.

In the last few years a number of different algorithms have been developed to search for monochromatic binaries.  A full list of these algorithms is given by~\cite{MLDC1, MLDC2, MLDC1B}, but we will mention a few of the more successful methods here :  a number of refined template grid methods have been used that use the F-Statistic to minimize the number of search parameters.  One method developed by Kr\'olak \& Blaut used the optimal placement of templates on a hypercubic lattice~\cite{MLDC1, MLDC2, MLDC1B}.  Another algorithm developed by Prix \& Whelan used a hierarchical method that searched for enforced trigger coincidences between TDI variables, followed by a coherent search using noise orthogonal TDI combinations~\cite{PW07,PWK08}.  Crowder, Cornish \& Reddinger developed a genetic algorithm where the different parameter set combinations for the GW waveform are treated as organisms in the parameter space and are allowed to ``evolve" by means of various biological rules~\cite{CCR06}.  

However, the most successful algorithms developed thus far (for all sources), have been based on Markov chain Monte Carlo (MCMC) methods.  Due to the success of these algorithms we describe in a little more detail the mechanics of MCMC methods.  

\subsection{Metropolis-Hastings \& Markov Chain Monte Carlo.}
The MCMC is a stochastic method which is ideal for searching through high dimensional spaces.  It works by constructing a chain of solution points in parameter space drawn from a proposal distribution that we believe to be close to the target density we are trying to model.  If the chain is run long enough then we are guaranteed to eventually map out the target density.  There are a number of types of MCMC.  The most popular being a Gibbs algorithm (where we update one parameter at a time) or a Metropolis-Hastings algorithm (where we update all parameters at the same time).  A Gibbs Markov chain is very simple to get up and running as it requires no a priori knowledge of the system.  However, as all proposals are accepted, it is possible to end up random-walking through the parameter space.  Also, Gibbs chains can get stuck on likelihood peaks if there is almost perfect (anti)correlation between parameters, or if our solution represents a single high peak in an otherwise featureless likelihood space.

The Metropolis-Hastings sampling method is a variant on the Markov Chain Monte Carlo method , and works as follows : starting with data $s(t)$ and some initial template $h(t, \vec{x})$, where $\vec{x}$ is a starting random combination of the system parameters,  we then draw from a proposal distribution and propose a jump to another point in the space $\vec{y}$.  In order to compare both points, we evaluate the Metropolis-Hastings ratio
\begin{equation}
H = \frac{\pi(\vec{y})p(s|\vec{y})q(\vec{x}|\vec{y})}{\pi(\vec{x})p(s|\vec{x})q(\vec{y}|\vec{x})}.
\end{equation}
Here $\pi(\vec{x})$ are the priors of the parameters,  $q(\vec{x}|\vec{y})$ is the proposal distribution  and $p(s|\vec{x})$ is the likelihood defined by
\begin{equation}\label{eqn:likelihood}
p(s|\vec{x}) = C\,e^{-\left<s-h\left(\vec{x}\right)|s-h\left(\vec{x}\right)\right>/2},  
\end{equation}
where $C$ is a normalization constant. This jump is then accepted with probability $\alpha = min(1,H)$, otherwise the chain stays at $\vec{x}$.  While the Metropolis-Hastings algorithm requires more effort, in that we need to tailor the proposal distributions to the problem at hand, it has a much faster convergence rate than Gibbs sampling.  Another way to improve convergence is to ensure that we are jumping along eigen-directions rather than coordinate directions.  This is achieved by using the eigenvalues and eigenvectors of the Fisher information matrix (FIM), i.e.   
\begin{equation}
\Gamma_{\mu\nu} = \left< \partial_{\mu}h | \partial_{\nu}h\right> ,
\end{equation}
to provide the scale and directionality of the jump.

At present a number of flavours of MCMC are being used within the GW community.  Some of these involve reverse jump MCMC~\cite{CL07, LC09} (where one can change the dimensionality of the search within the algorithm.  This is useful if we are also trying to carry out a model selection in terms of source number), delayed rejection MCMC~\cite{TVV09a, TVV09b} (here one proposes a move to a new point.  If the point is not accepted, instead of rejecting it straight away, it is kept and information from this point is used to aid proposal of another possible solution) and parallel tempered MCMC~\cite{LC09, KC09} (here, instead of running one chain, a number of cross-communicating chains of different temperatures are run simultaneously, allowing wider exploration the the hotter chains, and local exploration with the cold chains).

The MLDCs have become a way of testing algorithms within the gravitational wave community, especially in the case of galactic binaries.  The most important challenge thus far, in terms of displaying the capabilities of various algorithms, was MLDC 2.   In this challenge the groups were faced with two data sets.  One with a bare galaxy of approximately 30 million individually modelled galactic binaries, and a second set with again approximately 30 million galactic binaries, but also with between 4-6 SMBHBs in the data stream.    Of the groups that returned full parameter sets (i.e. all seven required parameters), the template grid of Kr\'olak \& Blaut found 404 sources in the bare galaxy challenge.  The hierarchical grid search developed by Prix \& Whelan did somewhat better with 1777 sources detected in the bare galaxy test, and 1737 sources in the galaxy plus SMBHB test.  However, by far, the most successful algorithm for the detection and extraction of parameters for galactic binaries was the Block Annealed Metropolis-Hastings or {\em BAM} algorithm developed by Crowder \& Cornish~\cite{CC04, CC05, CC07a,CC07b}.  

\subsection{BAM Algorithm.}
The BAM algorithm works by dividing the frequency search band into equal sized blocks.  The algorithm then sequentially steps through the blocks, simultaneously updating the sources within the blocks.  Once all the blocks are updated, they are shifted by one half a block width for the next round.  By breaking the band into blocks, fewer search templates are needed as compared to searching the entire frequency band at once.

The main issue, as always, with breaking up data streams, is the treatment of edge effects.  However, the BAM algorithm circumvents the problems in a number of ways.  Firstly, in each search region, two concepts entitled ``wings" and ``acceptance window" are introduced.  The acceptance window is, in effect, a safe region where one can be sure that search templates are finding reliable sources.  The wings have the job of dealing with the effects of the regions between the edge of the acceptance window and the boundaries of the search region.  One of their main jobs is to lessen the effect of bright sources that exist outside the search area and are bleeding power into the region of interest.  These sources cause an effect called ``slamming" where search templates are drawn to the boundary of the search area due to the bleeding power.  This can result in the templates missing other dimmer signals within the search area, and providing inaccurate estimations for the parameter values.  While these templates are recognizable due to large amplitudes and a high degree of (anti)correlation between the templates, they still represent a problem.   To overcome this problem, the authors introduced another concept called ``wing noise".  This is an exponential increase in the noise spectral density at the edge of the acceptance window.  This noise weights matches between templates and a signal less in the wings than in the acceptance window, thus encouraging templates to stay within the acceptance window.

\begin{figure}[t]
\begin{center}
\epsfig{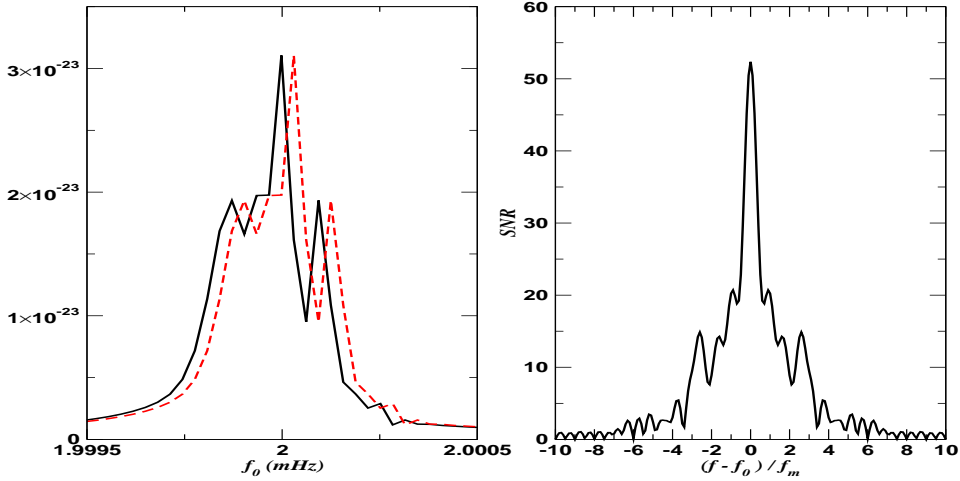}
\end{center}
\caption{On the left, we plot the same galactic binary at frequencies of $f_0 = 2$ and $f_0 = 2 + f_m$ mHz.  On the right, we plot the SNR as a function of frequency modulation offsets from $-10 \leq (f - f_0)/f_m \leq 10$.  We can see that an offset of $\pm 1 f_m$ produces secondary peaks in the SNR.  In general these, and other peaks, are separated at approximately $\pm n f_m$, where $n$ is an integer number.}
\label{fig:fmoffset}
\end{figure}

A number of techniques were used to increase the speed of convergence of the algorithm.  The first was an exploitation in certain symmetries which exist within the problem and the construction of corresponding proposal distributions.  This first is the fact that there is an antipodal sky solution due to the LISA response at approximately
\begin{equation}
\theta\rightarrow\pi-\theta,\,\,\,\,\,\,\,  \phi\rightarrow\phi\pm\pi.
\end{equation}
This symmetry allows one to force the algorithm to explore both hemispheres.  A second more subtle symmetry can be seen by Figure~\ref{fig:fmoffset}.  On the left hand side of this figure, we plot the power spectrum of two galactic binaries that are identical, except that the frequency of one is shifted by one LISA modulation frequency $f_m$.  On the right hand side of the figure, we plot the SNR obtained by keeping all parameters constant, except for the frequency of the template which we slide by $\pm10 f_m$.  We can see that there are symmetric peaks at distances of approximately 1$f_m$ separation across the band.  The authors used this information to ``island hop" through frequency space.  One of the other non-Markovian steps used by the authors was, if one island hop is accepted, immediately try another in order to converge as quickly as possible.

One of the issues with the convergence of any MCMC algorithm is the tendency for a chain to stay at a local maximum for a long time before moving on.  Simulated annealing          is a common way of softening features on the likelihood surface which causes a chain to get stuck, and thus allow wider exploration and faster convergence.   This is implemented by multiplying the noise weighted inner product by an inverse temperature
\begin{equation}
\beta = \frac{1}{T} = \left\{ \begin{array}{ll} \beta_0\left(\frac{1}{\beta_0}\right)^{i/Nc} & 0\leq i\leq N_{c} \\ \\ 1 & i > N_{c}  \end{array}\right.,
\end{equation}
where $\beta_0$ is the heat-index defining the initial heat, $i$ is the number of steps in the chain and $N_{c}$ is the cooling schedule.  Normally, simulated annealing is used to accelerate the chain to a stationary solution, after which the heat is set to unity and a fully Markovian chain is used.    It is also useful as it allows a chain to explore the likelihood surface faster as it is unlikely to get stuck on a secondary maximum while the heat is high.  The hope is that as we cool to unit temperature, the chain is already close enough to the true solution that it only has to walk uphill to the top of the central peak.  Some of the issues with simulated annealing include the fact there is no a priori information on what the value of the initial temperature should be.  If it is too high, we kill all features on the likelihood surface and waste a large number of computer cycles conducting a random walk in parameter space.  If it is too low, we very quickly find a local maximum and never move again within the timescale of the chain.  Also, with simulated annealing, one needs to cool the likelihood surface slowly, as too quick a cooling can again trap a chain on a local maximum.

\begin{figure}[t]
\begin{center}
\epsfig{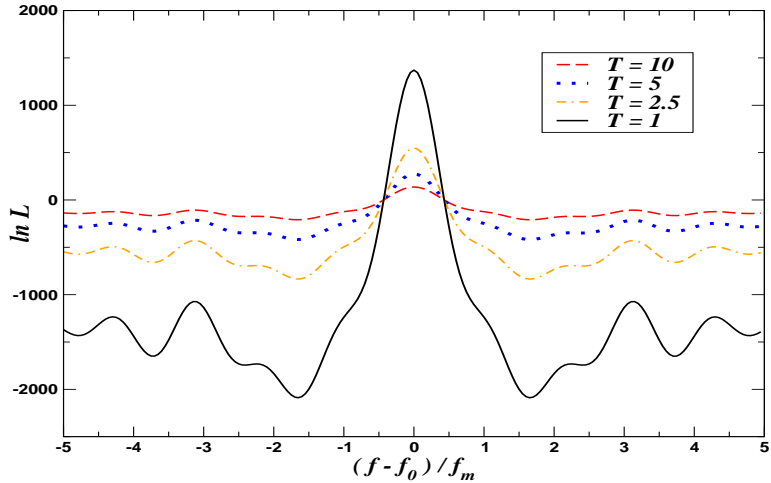}
\end{center}
\caption{A plot of the log-likelihood at frequency modulation offsets from $-5 \leq (f - f_0)/f_m \leq 5$ with different simulated annealing temperatures.  We can see that a heat of 10 essentially flattens all peaks on the likelihood surface apart from the main peak.  As we cool the surface, we begin to see the other peaks arising.}
\label{fig:sa}
\end{figure}

In Figure~\ref{fig:sa} we plot the log likelihood for a template across a frequency band of $\pm 5 f_m$ for four different values of the temperature $T$.  We can see that for $T=10$, only feature visible is the central peak.  However, as we cool from $T=10$ to 1, we can see the peaks in the likelihood surface becoming visible again.  Simulated annealing allows the chain to visit regions it may not otherwise go to.  For example, in the figure, we can see that a uniform temperature chain at $\pm 3 f_m$ is unlikely to move to the central peak in any reasonable amount of time due to the deep minima at $\pm 1.5 f_m$.  While a MCMC can move to areas of lower likelihood, it doesn't happen very often.  However, it is clear that these minima are not an issue for a chain with temperature $T=10$ and are easily traversable.  

In the previously mentioned MLDC 2, the {\em BAM} algorithm performed extremely well.  In the bare galaxy test, 19,324 sources were recovered, while in the galaxy plus SMBHB test, 18,461 sources were found.  These numbers are approaching the theoretical estimate for the number of recoverable sources.  Furthermore, the algorithm is very fast taking less than two weeks to process the entire galaxy using a 128 node cluster.

\subsection{Outstanding Issues for Galactic Binaries.}
It has become clear from the full galaxy challenges of the MLDC that there are two important questions to be answered.  While the questions are from different points of view, they are inextricably linked : what does one use as a termination criterion and how exactly does one evaluate many thousand possible sollutions simultaneously?  The common factor linking both questions is that we have no idea just how many sources will actually be resolvable, or what their parameter values will be.  On top of this, it is very difficult to even define the notions of false positives and false dismissals.  Some of the algorithms available today only return the intrinsic parameters, while others return all system parameters.  However, it is always possible to to get a better fit to a model with more parameters, so there may be cases where some algorithms are overfitting the data.  A number of groups have started to investigate Bayesian model selection for GW sources~\cite{CL07, LC09, VV08a, VV08b}, but the works are too detailed to discuss here.

\section{Non-Spinning Supermassive Black Hole Binaries.}
In this section we focus on algorithms for non-spinning SMBHBs.  A number of algorithms have now been developed that can successfully search for and extract the parameters from these inspiralling binaries.  In virtually all cases, the work has focused on restricted post-Newtonian binaries (i.e. while the phase is evolved to 2-PN order, the amplitude is kept at the dominant Newtonian level).  The SMBHB polarizations are given by
\begin{eqnarray}
h_{+}& =& \frac{2Gm\eta}{c^{2}D_{L}}\left(1+\cos^{2}\iota\right)x\cos(\Phi),\\ \nonumber \\
h_{\times} &= &-\frac{4Gm\eta}{c^{2}D_{L}}\cos\iota\,x\sin(\Phi).
\end{eqnarray}
Here $D_L$ is the luminosity distance from the source to detector.  The invariant PN velocity parameter is defined by $x = \left(Gm\omega / c^{3}\right)^{2/3}$, where 
\begin{eqnarray}\label{eqn:freq}
\omega(t)&=&\frac{c^{3}}{8Gm}\left[\Theta^{-3/8}+\left(\frac{743}{2688}+\frac{11}{32}\eta\right)\Theta^{-5/8}-\frac{3\pi}{10}\Theta^{-3/4}\right.\nonumber\\ &+&\left.\left(\frac{1855099}{14450688}+\frac{56975}{258048}\eta+\frac{371}{2048}\eta^{2}\right)\Theta^{-7/8}\right]\nonumber\\
\end{eqnarray}
is the 2 PN order orbital frequency for a circular orbit formally defined as $\omega=d\Phi_{orb}/dt$, and $\Phi =\varphi_{c}-\varphi(t) = 2\Phi_{orb}$ is the gravitational wave phase which is defined as
\begin{eqnarray}\label{eqn:phase}
\Phi(t) &=& \varphi_{c}-\frac{2}{\eta}\left[\Theta^{5/8}+\left(\frac{3715}{8064}+\frac{55}{96}\eta\right)\Theta^{3/8}-\frac{3\pi}{4}\Theta^{1/4}\right.\nonumber\\ &+&\left.\left(\frac{9275495}{14450688}+\frac{284875}{258048}\eta+\frac{1855}{2048}\eta^{2}\right)\Theta^{1/8}\right].\nonumber\\
\end{eqnarray}
The quantity $\Theta(t;t_{c})$ is related to the time to coalescence of the wave, $t_{c}$, by
\begin{equation}
\Theta(t;t_{c}) = \frac{c^{3}\eta}{5Gm}\left(t_{c}-t\right).
\end{equation}
In the search for SMBHBs, groups have used algorithms based on stochastic template banks~\cite{FFS09, BAB08}, time-frequency tracks~\cite{MLDC2}, user-refined grids~\cite{MLDC1} and a three step method combining time-frequency analysis, a template grid and a Metropolis-Hastings based Markov chain~\cite{BCCMV07}.  However, we present here three methods that have excelled in terms of speed and precision.

\begin{figure}[t]
  \vspace{5pt}

  \centerline{\hbox{ \hspace{0.0in} 
    \epsfxsize=2.6in
   \epsffile{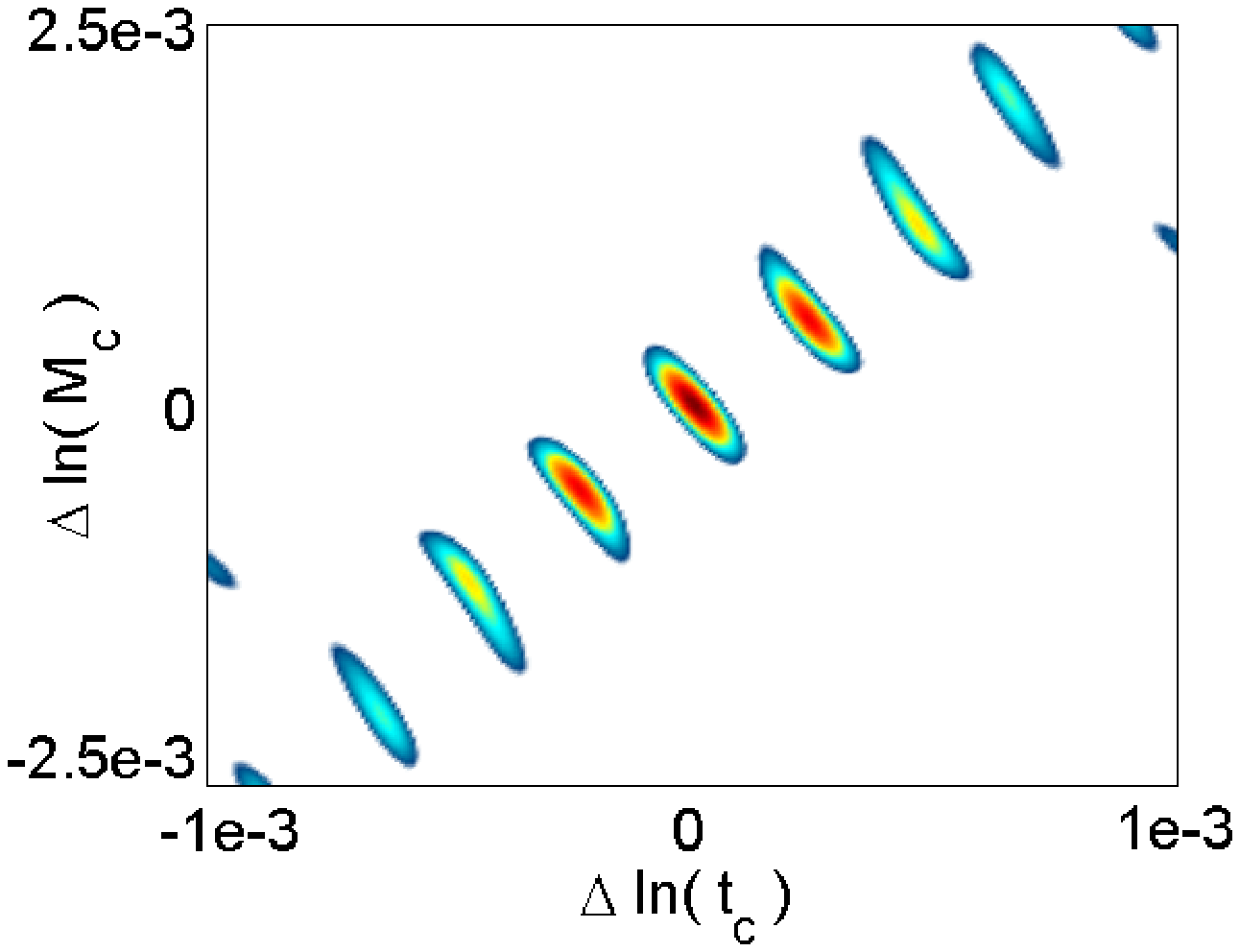}
    \hspace{0.05cm}
    \epsfxsize=2.7in
    \epsffile{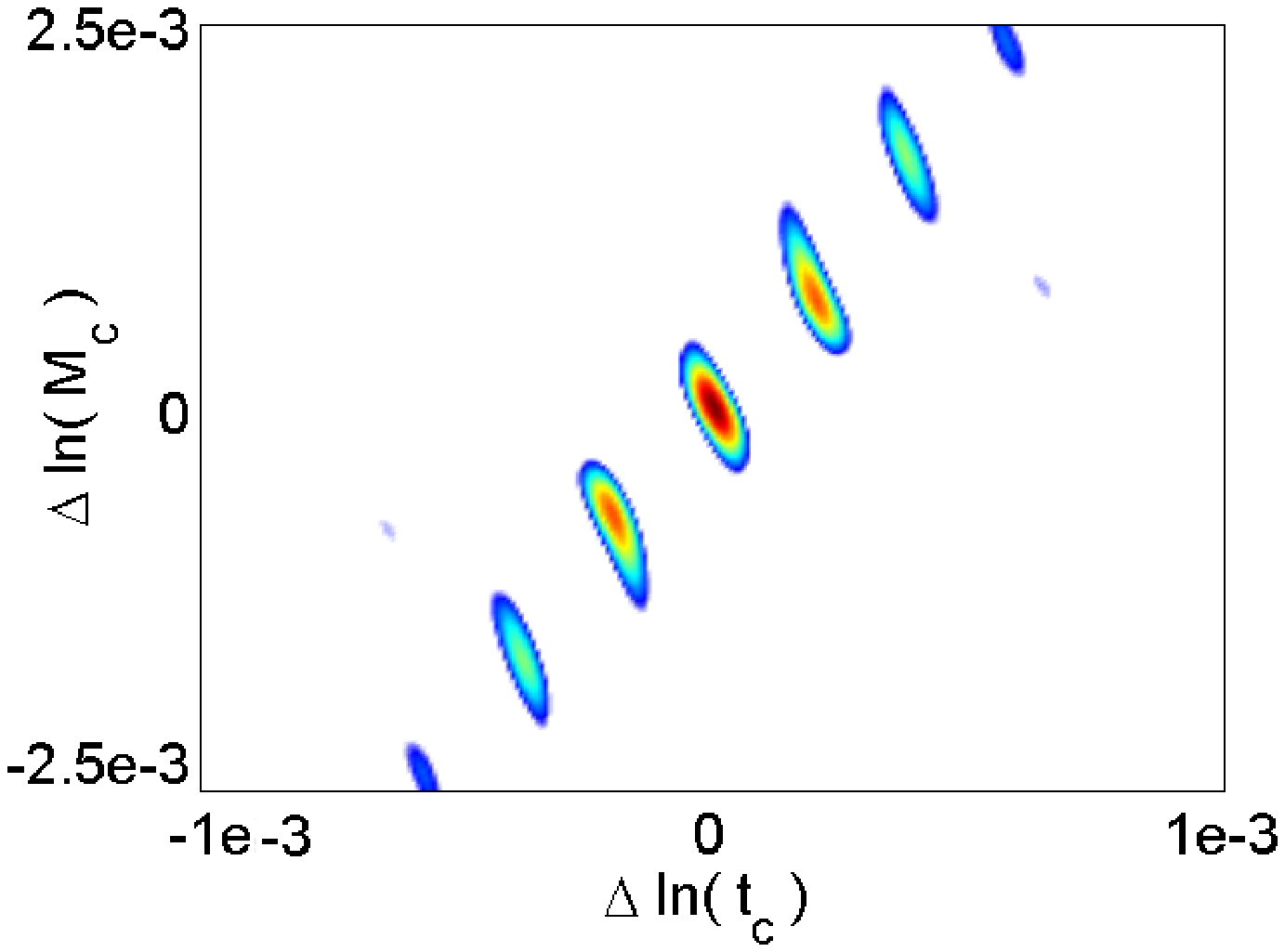}
    }
  }

  \vspace{5pt}
  \hbox{\hspace{1.in} ($t_c=T_{obs} +$ 1 month) \hspace{1.in} ($t_c=T_{obs} +$2 weeks)}   \vspace{7pt}

  \centerline{\hbox{ \hspace{0.0in}
    \epsfxsize=2.7in
    \epsffile{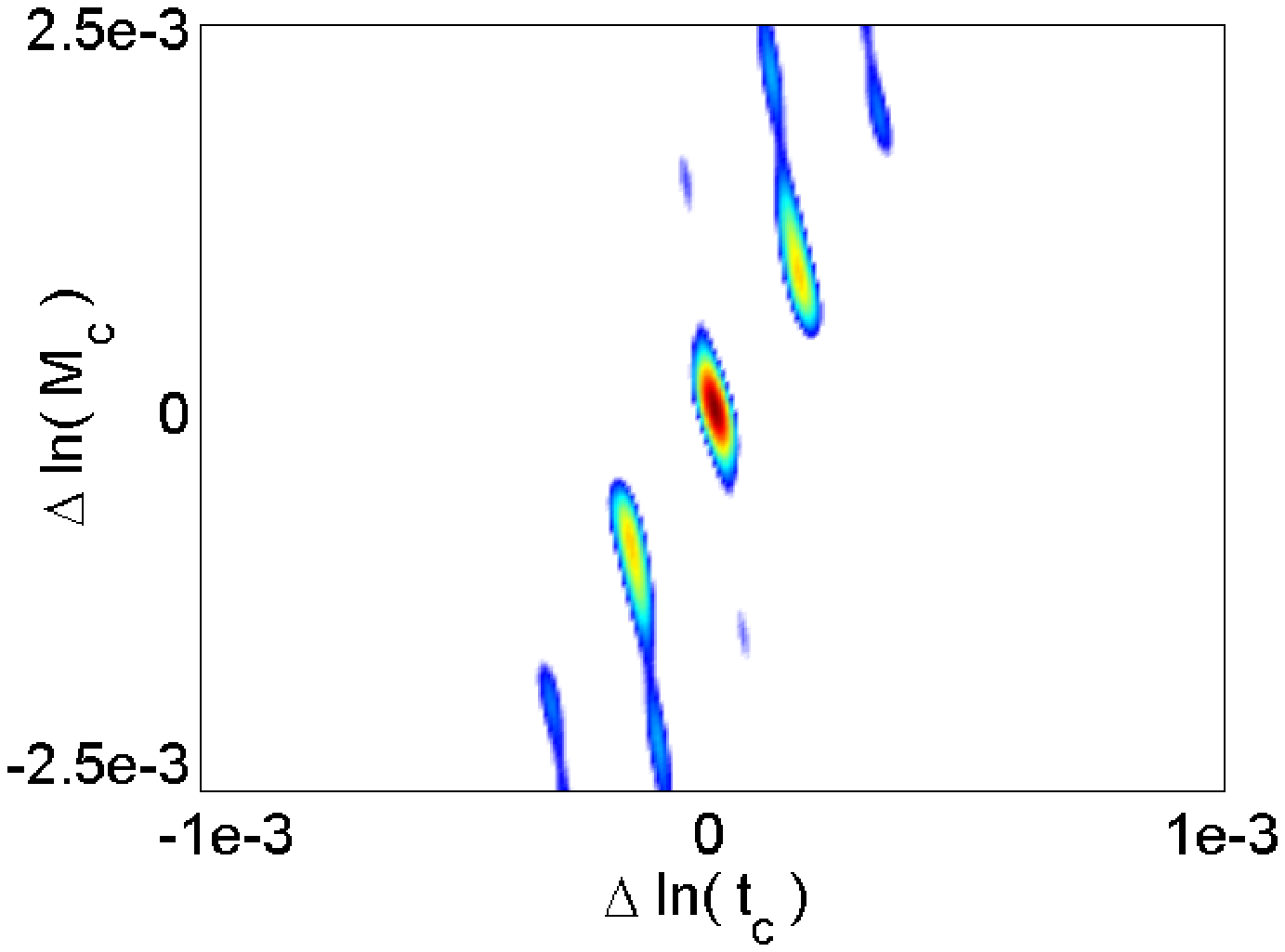}
    \hspace{0.05cm}
    \epsfxsize=2.7in
    \epsffile{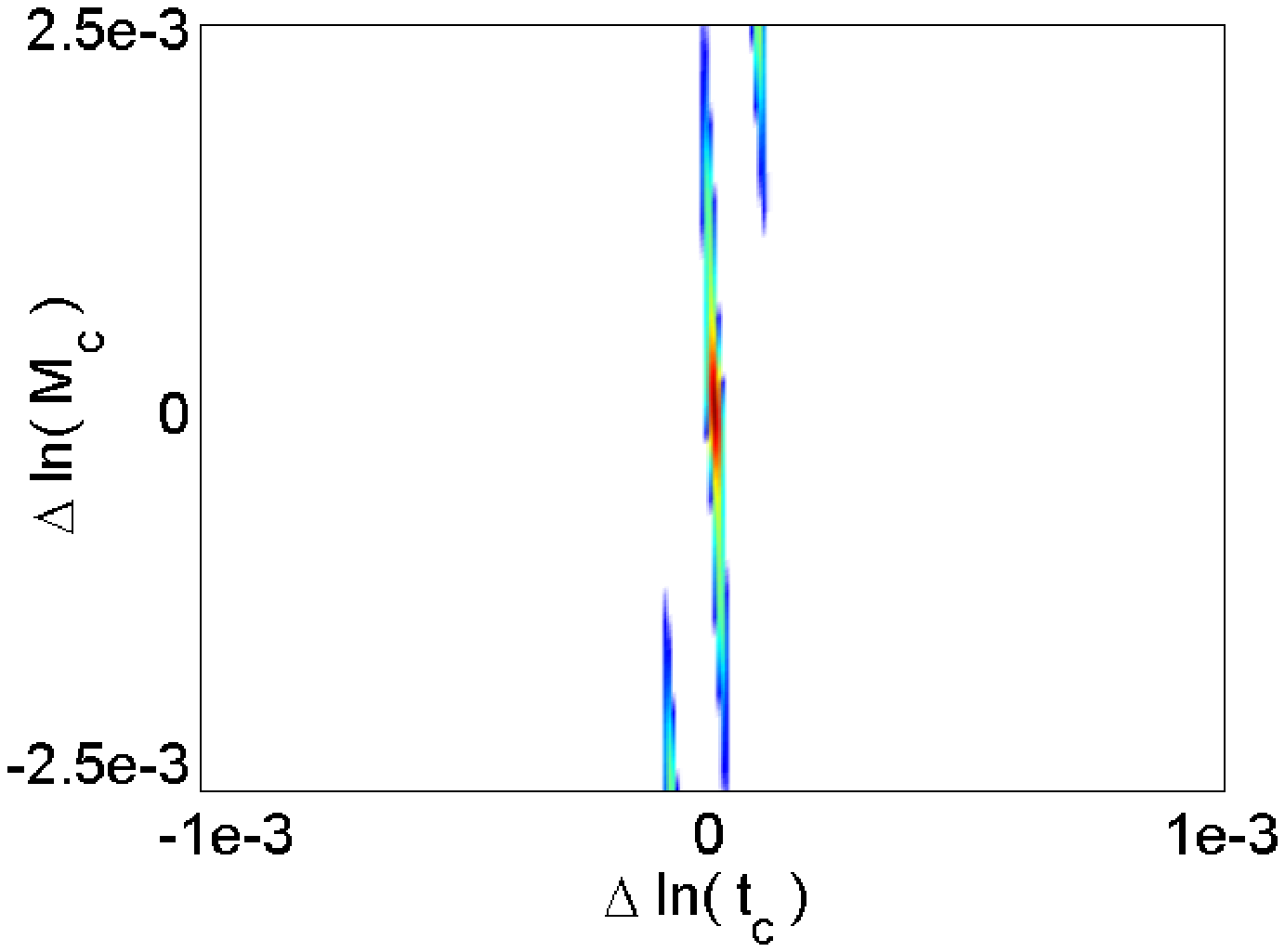}
    }
  }

  \vspace{5pt}
  \hbox{\hspace{1.in} ($t_c=T_{obs} +$1 week) \hspace{1.in} ($t_c=T_{obs} +$1 day)} 
  \vspace{5pt}

  \caption{ This Figure displays the evolution for the island chain of maxima on a $\ln\left(M_{c}\right)-\ln\left(t_{c}\right)$ slice through the Likelihood surface for coalescence times that exceed the observation time.}
  \label{fig:islandchains}

\end{figure}

\subsection{Metropolis-Hastings Search Algorithm.}
The Metropolis-Hastings search algorithm (MHMC) was developed by Cornish \& Porter~\cite{CP06, CP07a, CP07b, CP07c}.  Similar to the BAM algorithm, it used a number of tailored proposal distributions and techniques to speed up convergence of the waveform.  The first method of acceleration was that a version of the F-Statistic exists for non-spinning SMBHBs~\cite{CP07b}.  This allowed us to separate the parameter set into intrinsic $\{\ln M_c, \ln\mu, \ln t_c, \theta, \phi \}$ and extrinsic $\{\iota, \psi, \varphi_c, \ln D_L\}$ subsets, thus reducing the dimensionality of the search space form nine to five.

The second measure used was to accelerate the convergence of the algorithm by maximizing over the time-of-coalescence during the search and cooling phase of the chain.  For these sources,  this parameter is the most important one to find.  Once $t_{c}$ is found, due to the fact that the parameters are highly correlated, the mass parameters are usually found very soon after.  Strictly, this is not a correct step to use as it assumes that LISA is stationary.  However, during the annealing phase $t_{c}$ is treated as being a quasi-extrinsic parameter and searched over separately.  Once the cooling phase has finished this maximization is stopped.   The advantage of including this step in the search is that it manages to tie $t_{c}$ down to a restricted search range very quickly.  The $t_{c}$ maximization is carried out using a modified F-Statistic search.  Using the usual Fourier domain $t_{c}$ maximization, the equations for the F-Statistic take on a slightly different form.  This time instead of defining the four constants $N_{i}$, we define the matrix
\begin{equation}
N^{ij}=4\sum_{i=1}^{n}\sum_{j =1}^{4} \left(N^{ij}_{I}+N_{II}^{ij}\right),
\end{equation}
where $n$ is the number of elements in the waveform array.  The above equation describes the four constants at different time lags.  We can now solve for the time independent amplitudes
\begin{equation}
a_{ij} = \sum_{i=1}^{n}\sum_{j =1}^{4}\sum_{k =1}^{4}M_{jk}N^{ik}
\end{equation}
where the M-Matrix is the same as the one defined previously.  Inverting the M-Matrix, the F-Statistic is now a vector over different time lags.
\begin{equation}
{\mathcal F}_{i} = \sum_{i=1}^{n}\sum_{j =1}^{4}\sum_{k =1}^{4}M_{jk}N^{ij}N^{ik}
\end{equation}
This vector array is then searched through to find the value of $t_{c}$ that maximizes the F-Statistic.  While it is possible to carry out a search without this maximization, it takes much longer to converge.  In fact, with $t_{c}$ maximization the algorithm has found the time to coalescence of the SMBHB in as little as 10 steps of the chain for certain sources.  

A lot of the success of this algorithm was due to exploiting symmetries in the detector response, as well as features on the likelihood surface.  As in the case of galactic binaries, proposal distributions were used that forced the chain to explore the antipodal sky solution.   It was also found that, similar to the galactic binaries, ``island chains" in the likelihood surface could be exploited.  In Figure~\ref{fig:islandchains} we plot a slice through the $\ln\left(M_{c}\right)-\ln\left(t_{c}\right)$ surface.  We can see how the islands elongate and the chains rotate as we approach coalescence.  While the authors were unable to precisely work out the distance between islands, a proposal distribution was developed that allowed them to successfully exploit these island chains. 

One of the outcomes from the study of the island chains was that truncating the waveform before the coalescence changes both the shape, position and number of islands within a certain distance of the true solution.  We can see from Figure~\ref{fig:islandchains} that movement along the islands is a lot easier the further we are out from coalescence.  This, and the fact that it is very difficult to fit parameters for the most relativistic cycles at the end of the waveform, led the authors to develop an acceleration technique called {\em frequency annealing}.  With this scheme, the waveform at a particular iteration is generated to a cut-off frequency, $f_{cut}$, which is less than the maximum search bandwidth frequency, $f_{max}$.  This maximum bandwidth frequency is determined as a function of the lowest total mass in the priors.  The initial upper cut-off frequency is chosen to be a multiple (at least 2, but in most cases 4) of the lower frequency cutoff of LISA.  Then, by defining a growth parameter
\begin{equation}
B = \log\left(\frac{f_{max}}{a\,f_{cut}}\right)\,\,\,\,\,\,\,a\geq2, 
\end{equation}
we evolve the upper cut-off frequency according to 
\begin{equation}
f_{cut}= \left\{ \begin{array}{ll} 10^{-B\left(1-\frac{i}{N_{c}}\right)}f_{max} & f< f_{max} \\ \\ f_{max} & f \geq f_{max}  \end{array}\right.,
\end{equation}
This annealing scheme allows us to fit the early less relativistic cycles easily, thus constraining the parameters.  In practice, as we approach the point where we are generating full templates, the waveform parameters are already close to the correct values.  This is due to the fact that terminating the templates early manages to surpress a lot of the features on the likelihood surface that would normally present an obstacle to the chain.  In this respect, the frequency annealing acts as a form of simulated annealing as it changes the structure of the likelihood surface.  One of the main advantages of this scheme is that the first few thousand iterations of the chain are achieved in less than ten minutes.  This allows us to give a very quick confirmation of a detection.

The final refinement used was to introduce a {\em thermostated} heat factor.  As the frequency annealing is a progressive algorithm (i.e. it does not see the full signal until very late in the run) it is important the algorithm does not get stuck on a secondary maximum.  To get around the authors used a thermostated heat of 
\begin{equation}
\delta = \left\{ \begin{array}{ll} 1.0 & 0\leq SNR\leq SNR_0 \\ \\ \left(\frac{SNR}{SNR_0}\right)^{2} & SNR > SNR_0  \end{array}\right. ,
\end{equation}
which ensures that once a SNR of greater than $SNR_0$ is attained, the effective SNR never exceeds this value (this value is chosen a priori by the user, but it was found that setting $SNR_0=20$ usually suffices).  This thermostated heat means that the chain explores the parameter space more aggressively and as the algorithm approaches using the full templates, there is enough heat in the system to prevent the chain from getting stuck.  This stage is carried out for a certain number of iterations and then cooled using standard simulated annealing.

One other thing worth mentioning here is the ability of this algorithm to also map out the posterior density functions (pdfs) for each solution.  It was shown that on the projected 5-D parameter space of the intrinsic parameters, the pdfs obtained from the MCMC were a very good match to the predictions of the FIM.  However, there were cases where the MCMC displayed a deviation from the predictions of the FIM due to high correlations between the parameters.  This showed that, while in many cases we can take the estimations of the FIM as a good approximation, there are sometimes when it (under)overestimates the error.  We refer the reader to a paper by Vallisneri~\cite{VAL08} on the validity of the FIM in GW astronomy.  The authors also showed that when trying to map out the pdfs for sources where we do not see coalescence, the large uncertainty in the phase at coalescence $\varphi_c$ causes the chains to go on a random walk, and the mapping of the pdf takes a very long time.  To circumvent this, the authors used a {\em mini F-Statistic} where only the luminosity distance and phase at coalescence are maximized over.  Expanding the detector response in terms of $\varphi$ and $\varphi_{c}$ we obtain
\begin{eqnarray}
h(t) &=& \frac{\cos(\varphi_{c})}{D_{L}}\left[A_{+}F^{+}\cos(\varphi)-A_{\times}F^{\times}\sin(\varphi)\right]\nonumber\\
&+&\frac{\sin(\varphi_{c})}{D_{L}}\left[A_{+}F^{+}\sin(\varphi)+A_{\times}F^{\times}\cos(\varphi)\right].
\end{eqnarray}
The square bracket terms in the above expression correspond to the responses $h(t;\varphi_{c}=0)$ and $h(t;\varphi_{c}=\pi/2)$ respectively, with the luminosity distance set to unity, which allows us to write the mini F-Statistic in the form
\begin{eqnarray}
h(t) & =&  \frac{\cos(\varphi_{c})}{D_{L}}\,h(t; \varphi_{c}=0, D_{L}=1)
 + \frac{\sin(\varphi_{c})}{D_{L}}\,h(t;\varphi_{c}=\pi/2, D_{L}=1)\nonumber\\
     & = & \sum_{k=1}^{2}\,a_{k}A^{k},
\end{eqnarray}
where
\begin{equation}
a_{1} = \frac{\cos(\varphi_{c})}{D_{L}}\,\,\,\,\,\, , \,\,\,\,\,\,\,\,\,
a_{2} = \frac{\sin(\varphi_{c})}{D_{L}},
\end{equation}
and
\begin{equation}
A^{1} = h(t; \varphi_{c}=0, D_{L}=1)\,\,\,\,\,\, , \,\,\,\,\,\,\,\,\, A^{2} = h(t;\varphi_{c}=\pi/2, D_{L}=1).
\end{equation}
Repeating the steps from the generalized F-Statistic, one can obtain the numerical values for the quantities $a_{k}$. The maximized values of $\varphi_{c}$ and $D_{L}$ are then found using the expressions
\begin{equation}
\varphi_{c} = \arctan\left(\frac{a_{2}}{a_{1}}\right),
\end{equation}
\begin{equation}
D_{L} = \left[a_{1}^{2} + a_{2}^{2}\right]^{-1/2}.
\end{equation}
The above process greatly improved the convergence of the chains for the non-coalescing sources.

In both user trials and blind MLDC challenges, this algorithm performed exceptionally well.  Parameters were always estimated to within a 5$\sigma$ error.  However, the main attraction of the algorithm was the fact that the total run-time was 4-5 hours on a laptop.  While one would always run multiple chains to ensure detection and parameter estimation, this algorithm does not require the use of a cluster and is thus easily accessible to others.

\subsection{Hybrid Evolutionary Algorithm.}
The MHMC algorithm described above is an iterative algorithm in that it first finds the brightest source, removes that source and then searches for another.  The Hybrid Evolutionary Algorithm (HEA) was developed by Gair \& Porter~\cite{GP09} in an effort to simultaneously find multiple mode solutions for multiple sources.  The algorithm uses a combination of Metropolis-Hasting, Nested Sampling and evolutionary rules to solve the problem.   The version of Metropolis-Hastings used is slightly different from the one described above in that the Metropolis-Hastings ratio is now decided by
\begin{equation}
H =  \left\{ \begin{array}{ll} 1 &{\mathcal L}(\Theta')>{\mathcal L}_{i}\,\, \mbox{and}\,\, \pi(\Theta')>\pi(\Theta) \\ \\ \pi(\Theta')/\pi(\Theta) & {\mathcal L}(\Theta')>{\mathcal L}_{i}\,\, \mbox{and}\,\, \pi(\Theta')\leq\pi(\Theta) \\ \\ 0 & \mbox{otherwise}  \end{array}\right. ,
\end{equation}
where again $\pi(\Theta)$ are the priors based on the parameters $\Theta$, and ${\mathcal L}(\Theta)$ denote the likelihood.  Nested Sampling was introduced by Skilling~\cite{SKI04} as a tool for evaluating the Bayesian evidence, by employing a set of live points that climb together through nested contours of increasing likelihood.  At each step, the algorithm attempts to find a point with a likelihood higher than the lowest likelihood point in the live point set and then replaces the lowest likelihood point with the new point. Nested sampling has already been applied to the issue of model selection for ground based observations of gravitational waves~\cite{VV08b}

The HEA algorithm works as follows : a number of organisms (trial solutions) and a predefined number of cluster centroids are dropped onto the likelihood surface.  The initial organisms must satisfy a fitness criterion to survive (i.e. their initial SNR must be greater than some threshold).  After that each successive organism is required be fitter than the mean fitness of the group.  Once the initial selection has been made, the organisms are given the chance to improve their fitness, before being allowed to evolve through the system by using the Metropolis-Hastings, Nested Sampling algorithm or by using rules from evolutionary computation such as birth, death, altruism etc.  At preset points the organisms are clustered according to the Euclidean distance from the center of mass of a cluster and either stay where they are, or join a new cluster of solutions.  The centroids, while initially dropped randomly, are then assigned to the center of mass of each cluster.  While we are interested in the evolution of each individual organism, it is the evolution of each centroid that we are truly interested in.  The HEA uses some of the same simulated and thermostated annealing concepts from the MHMC algorithm.  As each cluster needs to evolve at its own pace, we divide the parameter space into {\em Voronoi regions}.  This associates a certain amount of the parameter space with a particular cluster.  The temperature in each Voronoi region is then linked to the fittest member of each cluster.  This allows some clusters to evolve faster than others, but also means that the temperature associated with an extremely bright source does not kill the features associated with a dimmer source.

\begin{figure}[t]
\begin{center}
\includegraphics[width=\textwidth, keepaspectratio=true]{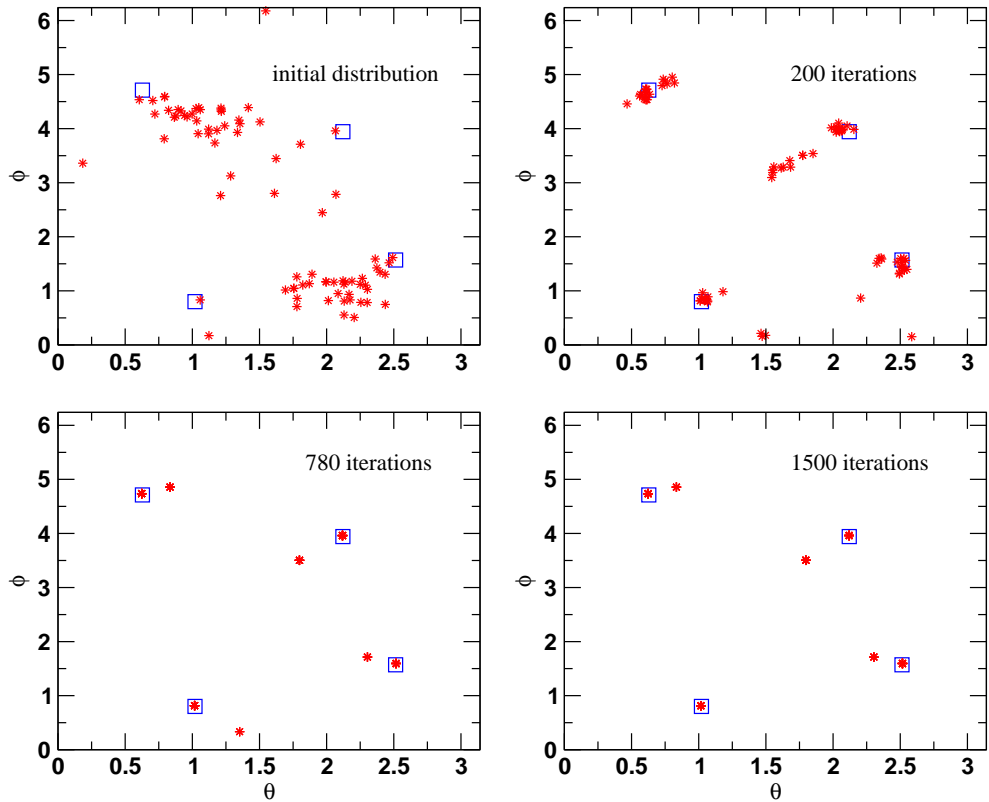}
\end{center}
\caption{A plot of the HEA sky search at four different times.  In each cell, the square represent the true solutions, while the stars represent the organisms.  We plot (going from top-left to bottom-right) the initial distribution after the initial selection and uphill climber improvement phase, and then at 200, 780 and 1500 iterations.  We see that not only does the algorithm find the two primary sky solutions (i.e., real and antipodal) for both sources, but also a bunch of secondary solutions at almost 90 degrees to the primaries.  } 
\label{fig:skysearch}
\end{figure}

The algorithm was tested on a data set with two SMBHBs.  While there were only two sources, things were complicated by the fact that there was a bright coalescing source, plus a dimmer non-coalescing source.  The algorithm was informed that there were between two and eight possible solutions and 80 live points were used in the search.  In Figure~\ref{fig:skysearch} we focus on the search over sky parameters (due to known degeneracies in the parameter space ) and plot the performance of the algorithm at four different snapshots.  We can see that the algorithm converges very quickly.  It not only finds the primary modes associated with each source (i.e. true and antipodal), but also manages to find a number secondary modes as well.  For both sources, the intrinsic parameters were all found to within 5$\sigma$ of the true values.  Just like the MHMC algorithm, the HEA again requires no more than a laptop to search for non-spinning sources, with a run time of about 10 hours.

\subsection{MultiNest.}
One of the problems with the Nested Sampling algorithm is, given a position in parameter space, the hard constraint of randomly finding a better point within the prior volume.  At the start of an algorithm when we are far away from the true solution, this is not that much of a constraint, but as we get closer to the true solution this becomes more and more difficult.  As a result, the acceptance rate of proposals starts to rapidly decrease, and the runtime begins to increase accordingly.  MultiNest is a multimodal nested sampling algorithm designed by Feroz, Hobson \& Bridges~\cite{FHB08} to efficiently evaluate the Bayesian evidence and return posterior probability densities for likelihood surfaces containing multiple secondary modes in Cosmology and particle physics.  It solves the problem of likelihood evaluation by using the current set of live points as a model of the shape of the likelihood surface. The algorithm uses an ellipsoidal rejection sampling scheme by enclosing the live point set into a set of (possibly overlapping)  ellipsoids and then uniformly draws a new point from the region enclosed by these ellipsoids.  This allows highly correlated pdfs to be broken into a number of smaller regions of overlapping ellipsoids.  The other main advantage of the MultiNest algorithm is that it allows both the local and global evidence to be evaluated.

MultiNest was recently applied to GW astronomy for non-spinning black holes by Feroz, Gair, Hobson \& Porter~\cite{FGHP09}.  The algorithm was run on the same two sources that were used for the HEA, using approximately 1000 live points.  The MultiNest algorithm is model independent in that it only requires calculation of the likelihood, and is therefore very fast.  In the initial part of the search, an F-Statistic was used to reduce the parameter space to five dimensions.  The algorithm found eleven modes in total, including the seven modes found by the HEA.  A second stage was also run, where the extrinsic parameters were also searched for.  In all cases the parameters were again found to with 5$\sigma$, and just like the other two algorithms described above, could be run on a laptop with a runtime of $\sim$ 3 hours.

\subsection{Outstanding Issues for SMBHBs.}
The binary black hole problem is much simpler than the galactic binary problem in that we do not expect to have very many SMBHBs in the data stream at any one time.  Even if we are faced with a number of simultaneous sources, it was shown by Cornish \& Porter~\cite{CP07b} that SMBHBs will be invisible to each other, so we will not have to contend with a confusion problem between sources.  At present, there are a number of end to end, or multiple stage algorithms that can detect and extract the parameter sets for both individual and multiple non-spinning massive black hole binaries.  However, in recent years it has been shown by Hughes \& Lang~\cite{LH06, LH08} that the inclusion of spin will have a dramatic improvement on the estimation of parameters.  More recently, it has been shown by a number of groups (see for example : Porter \& Cornish~\cite{PC08}, Trias \& Sintes~\cite{TS08a, TS08b}, Arun et al~\cite{AISS07}) that the corrections to the waveforms from higher harmonics break correlations between parameters and can also improve parameter estimation.  Therefore, at some point, algorithms that can search for spinning SMBHBs with higher harmonic corrections will be needed for LISA. While we have not covered it here, blind tests for spinning SMBHBs have already started in MLDC 3.  Finally, all existing algorithms search for the inspiral phase only.  In reality, we will need to develop algorithms that will search for signals that also include a merger and ringdown phase.  While the ringdown phase is quite easily modelled~\cite{BCC07}, we must wait until numerical relativity has increased in accuracy, sufficient for the development of LISA waveforms.

\section{Extreme Mass Ratio Inspirals.}
The inspiral of a stellar mass black hole or neutron star into a supermassive black hole can provide a richness of information due to the fact that the compact object spends a large fraction of its observable life in the highly relativistic region close to the massive black hole.  It has been suggested that due to the large number of observable GW cycles ($\sim10^5$), that EMRIs can be used to test General Relativity by mapping out the strong-field spacetime around the central black hole.  For a more detailed overview of EMRI astrophysics and detection with LISA, we refer the reader to an article by Amaro-Seoane et at~\cite{am07}.

However, due to the highly relativistic nature of the source, it is extremely difficult to model the compact object's orbit.  While advanced methods exist to approximate the inspiral, the generation time for these waveforms puts any data analysis effort currently out of reach.  In recent years the community has been using the analytical kludge waveforms of Barack \& Cutler~\cite{BC04}.  These are phenomenological waveforms that capture the complexity of true EMRI signals and importantly, have the correct number of paramters.  In general, this model is described by a 14-D parameter set : the mass of the central black hole $M$, the mass of the compact object $\mu$, the luminosity distance to the source $D_L$, the initial frequency $\nu_0$ or plunge time $t_p$, the initial(final) eccentricity $e_0 (e_p)$, the spin of the central black hole $S$, the position of the source in the sky $(\theta_S, \phi_S)$, orientation of the black hole spin $(\theta_k, \phi_k)$, inclination of orbit $\lambda$ and three initial orbital phases $(\Phi_0, \alpha_0, \gamma_0)$.

With modern computing abilities we are restricted to searching within a very narrow range of priors for EMRI sources.  This is due to a number of issues.  Presently, the waveform generation codes for the MLDCs take many minutes to generate a two year data set, as the waveforms contain many harmonics.  Even something as simple as calculating the likelihood becomes a time consuming chore with this waveform.  Things get  rapidly worse if we need to calculate the FIM, as we need to generate 28 waveforms to calculate the derivatives of the waveforms with respect to the parameters.  It is clear that this kind of time consumption is clearly unusable.  A lot of the recent effort has been made in efforts to speed up the generation of the waveforms.  This has involved restricting the number of waveform harmonics, approximating Bessel functions etc.  A waveform code developed by Cornish~\cite{COR08} is now orders of magnitude faster than the codes used for the MLDC data generation and makes data analysis a reality.  

It was shown by Gair et al~\cite{GBCCLPV04} that a reasonable template grid search for EMRIs is clearly out of reach as it would require approximately $10^{40}$ templates.  At present two approaches have been used to search for individual EMRIs buried in instrumental noise : a time-frequency analysis and a Metropolis-Hastings based algorithm.  We will look at each in detail later on.  As we said earlier, the convergence of the MH based algorithm can be greatly improved if we jump along eigen-directions rather than coordinate directions.  This requires frequent calculation of the FIM.  As well as the time constraints on the generation of the FIM, it was found that one has to be very careful with the numerical methods used for inverting the FIM.  The FIM for EMRIs is highly singular with a large condition number.  As the FIM is calculated and inverted numerically, it was found that while we achieved convergence for the elements of the FIM, the elements of the inverse could differ by orders of magnitude.  Alot of work has been done in the last year or so by Cornish, Gair \& Porter on the numerical stability of the FIM inversion as part of the LISA Parameter Estimation Taskforce (LPET\footnote[1]{http://www.tapir.caltech.edu/dokuwiki/lisape:home}).

There is still a lot of work to do for EMRI sources.  We are not yet in a situation where we have algorithms as developed as in the case of galactic binaries or non-spinning SMBHBs.  However, algorithms exist that can detect and carry out parameter estimation for isolated high SNR EMRIs embedded in instrumental noise.  We discuss the two current approaches below.

\subsection{Time-Frequency Methods.}
The time-frequency method for EMRI searches~\cite{WG05a,WG05b,WGC06, GJ07} works by dividing the data into a number of segments.  Each segment is Fourier transformed to produce a spectrogram of the data.  One then searches the spectrogram for features.  One of the first methods used was the Hierarchical Algorithm for Clusters and Ridges (HACR).  This algorithm was shown by Gair \& Jones~\cite{GJ07} to have a 10-15\% higher detection rate than a simple excess power algorithm.  However, after the MLDC 2, the authors improved the algorithm by introducing an EMRI tuned track search called the Chirp-based Algorithm for Track Searches (CATS)~\cite{GMW08a, GMW08b}.  In this algorithm, one constructs a 3-D grid in the the $(f, \dot{f}, \ddot{f})$ parameters space.  At each point, a potential track is constructed, and the power in each pixel along the track is added to give the total track power.  The brightest track is then claimed as a detection.  The power in the pixels of this track are then set to a large negative value to prevent intersection with future tracks.

In the MLDCs, the time-frequency algorithm worked quite well.  The only drawback with this type of search is that while cheap to use computationally, there is a corresponding loss in resolution.  The algorithm only returns the intrinsic parameters, so may only work as a first step algorithm.  Also, it is unclear how the algorithm will perform when we have overlapping sources plus a galaxy.

\subsection{Metropolis-Hastings Based Algorithms.}
To this point, two groups have developed MH based search algorithms for EMRIs~\cite{COR08, GPBB08, BGP09}.  The two algorithms are very close in function, due to having a common starting point in the BAM and MHMC algorithms.  The current problem with EMRI algorithms is the time it takes for development.  As the galactic binary and SMBHB waveforms were very quick to generate (i.e. many hundreds per second), the development of these algorithms was also quite rapid.  It could be seen in a matter of minutes/hours that there was a feature on the likelihood surface that was trapping the chain.  A proposal distribution could quickly be formulated to bypass this problem and the development continued.  With the EMRIs however, the situation is a lot more complicated.  Due to the frequency of the MLDCs, there hasn't been time to properly explore the EMRI likelihood surface.  Also, the chains take a long time to run, sometimes up to weeks at a time, so it is very hard to quickly see if there are obstacles to convergence.

However, progress has been made on this front.  As already mentioned, the current algorithms use the aforementioned concepts of simulated and thermostated annealing.   The main difference compared to other searches, is that the current algorithms are fundamentally hierarchical due to the time restrictions detailed above.  Shorter waveforms (3$\sim$4 months duration) are used to improve on the starting point in parameter space.  The best points from this initial stage are then taken as starting points for the next round, where the length of a template is possibly increased.  While a true F-Statistic is not available for EMRI waveforms, it is possible to optimize over the luminosity distance, the plunge time and the three initial phases using a combination of methods described in the SMBHB section above.  By matching individual harmonics, it is also possible to now conduct some "island hopping" in the parameter space.

Both algorithms have thus far been tested in situations where we have isolated EMRIs buried in instrumental noise.  In this case both algorithms have successfully detected and carried out parameter estimation for these sources.  However, we should mention once again, that due to computational constraints, the parameter priors are very narrow and it is impossible to extrapolate the performance of each algorithm to a wide prior search.  More recently, the algorithms have been applied to overlapping EMRIs in the third MLDC, so it will be interesting to see how each algorithm performs.

\subsection{Outstanding Issues for EMRIs.}
As said, we still have nowhere near the same comprehension of the likelihood surface for EMRIs that we have for other sources.  This requires serious study in order to further improve the current algorithms.  While the Barack-Cutler waveforms capture the complexity of EMRI orbits, at some point a transition to more exact numerical waveforms will have to be made.  It could be that a different waveform model could introduce (or remove) features on the likelihood surface that will dramatically effect the convergence of algorithm.  Finally, it is still not clear how the algorithms will perform once we widen the parameter priors or begin to have some EMRI confusion due to a high number of sources in the data stream.

\section{Conclusions}\label{sec:conclusions}
We have presented here an overview of the current state of the art algorithms for LISA data analysis.  While a number of different approaches have been used, we focused on those that were superior in terms of both accuracy and speed.  At present an algorithm exists for galactic binary analysis that is already approaching the theoretical limits for the resolution of binaries in the galaxy.  While the algorithm can be used on a desktop, it can process an entire galaxy of $\sim$30 million binaries in under two weeks on a cluster.  We also presented three detection and parameter estimation algorithms for non-spinning SMBHBs that currently require no more than a laptop to run on.  These algorithms provide very accurate parameter estimation and search over wide priors in less than half a day.  Finally, we introduced the currently available algorithms for EMRI data analysis.  This is the area with the least amount of development.  This is due to the complexity of the problem, the time taken to carry out the analysis and a not very advanced understanding of the likelihood surface for these sources.  Even so, the current algorithms are able to detect and extract the parameters for isolated EMRIs buried in instrumental noise.  We again caution the reader that it is too early to make predictions of the effectiveness of these algorithms as a lot more work is required.  

With approximately a decade to go before the planned LISA launch date, we are in a very good state concerning LISA data analysis.  There has been a significant progression from year to year in the methods and algorithms used to search for and detect various sources.  Lessons that have been learned from previous algorithms, have fed into the next generation of development, allowing more and more sophisticated techniques to be used.  While some methods have passed by the way, the good news is that new algorithms are always popping up.  And, while we still have a long way to go, given the current algorithms, we can proceed with a certain amount of confidence.

\section*{References in the highlight article}

\end{document}